Routledge
Taylor & Francis Group

**OPEN ACCESS**

Check for updates

# Governing autonomous vehicles: emerging responses for safety, liability, privacy, cybersecurity, and industry risks

Araz Taeihagh  and Hazel Si Min Lim

Lee Kuan Yew School of Public Policy, National University of Singapore, Singapore

**ABSTRACT**

The benefits of autonomous vehicles (AVs) are widely acknowledged, but there are concerns about the extent of these benefits and AV risks and unintended consequences. In this article, we first examine AVs and different categories of the technological risks associated with them. We then explore strategies that can be adopted to address these risks, and explore emerging responses by governments for addressing AV risks. Our analyses reveal that, thus far, governments have in most instances avoided stringent measures in order to promote AV developments and the majority of responses are non-binding and focus on creating councils or working groups to better explore AV implications. The US has been active in introducing legislations to address issues related to privacy and cybersecurity. The UK and Germany, in particular, have enacted laws to address liability issues; other countries mostly acknowledge these issues, but have yet to implement specific strategies. To address privacy and cybersecurity risks strategies ranging from introduction or amendment of non-AV specific legislation to creating working groups have been adopted. Much less attention has been paid to issues such as environmental and employment risks, although a few governments have begun programmes to retrain workers who might be negatively affected.

## Introduction

Autonomous vehicles (AVs) develop new paths for mobility and are acknowledged to have economic and societal benefits, but there are concerns regarding the extent of their benefits and their unintended consequences. As with all new technologies, appropriate governance strategies can help maximise the potential benefits associated with the rapid development of AVs and minimise the risks often associated with technological disruption and negative and/or unintended consequences. The concern, however, remains about the capacity of governments in the timely management of wider societal implications.

**CONTACT** Araz Taeihagh ✉ spparaz@nus.edu.sg, araz.taeihagh@new.oxon.org 🏛 Lee Kuan Yew School of Public Policy, National University of Singapore, 469B Bukit Timah Road, Li Ka Shing Building, Level 2, #02-10, Singapore 259771

Since Google released its first fleet of AVs in 2010 (Teoh & Kidd, 2017), developments in AV technology have accelerated significantly. Hillier, Wright, and Damen (2015) estimate that auto companies will roll out AVs in the market by 2020 and AVs are expected to occupy 25% of the global market by 2040 (West, 2016). Most scholarly work has been directed towards the effects of AVs. For instance, Milakis, Snelder, van Arem, Homem de Almeida Correia, and van Wee (2017) and Wadud, MacKenzie, and Leiby (2016) estimate the impact of AVs on transport demand and energy consumption respectively, while Collingwood (2017) and Glancy (2012) explore the impact of AVs on privacy issues.

There is limited country-specific literature regarding the policy implications of AVs and governance responses to AVs. Literature reviews have been conducted for Australia (Hillier et al., 2015; Sun, Olaru, Smith, Greaves, & Collins, 2016) and the United Kingdom (UK) (Clark, Parkhurst, & Ricci, 2016). Kalra (2017) has identified regulatory gaps in the United States (US) federal government's approach to AV-related safety risks, proposing possible risk management strategies. These studies, however, do not explicitly analyse government strategies and efforts as part of a broader framework. This article addresses the following questions: (a) what are the different kinds of risks associated with AVs? (b) what are the emerging government responses to address these risks and how can these different emerging strategies be categorised and compared? To address these two questions, we focus on efforts at a national level and consider the broader developments in the European Union (EU) as well.

The next section briefly introduces the methodology used for the selection of articles and reports on the governance of AVs. We provide the necessary background information about AVs before discussing the risks associated with them. We present a theoretical framework for examining responses to risk associated with AVs and identify and discuss the various emerging strategies applied by governments to address these risks before the concluding remarks.

## Methodology

Our methodology involved two steps. Firstly, we identified AV-related implications by preliminary review and exploration of the key factors that were highlighted as the most prominent in the current literature. We searched for possible risks associated with AVs using the keywords "autonomous vehicle(s)", "driverless" or "driverless vehicle(s)" in combination with one of the following keywords representing an AV-related implication (Table 1). Boolean operators such as "AND", "OR" and "NOT" were also used. To identify the lesser-known risks of AVs, we searched AVs in conjunction with "risk(s)" and its synonyms, such as "effect(s)", "impact(s)" and "consequence(s)". Secondly, existing government efforts to manage AV-related risks were identified. We searched for words relating to government regulation, such as "regulation(s)", "legislation(s)", "rule(s)", "bill(s)" and "law(s)", together with AVs and the names of the countries and regions of study. These include Australia, China, the EU, Germany, Japan, South Korea, Singapore, the US, and the UK, as most of AV-related developments have occurred in these regions and countries.

Research published from 2000 and onwards was obtained from well-known academic databases (Scopus, ScienceDirect, Web of Science and Springer). Google Scholar was used as the search engine only when the databases produced limited or no results for specific implications of AVs. News articles were also used to supplement background research on AVs, such as NYTimes, The Guardian, and Reuters. Government reports and policy documents were also included to identify the previous, current, and future government measures to address AV-related risks.

**Table 1.** Keywords used to identify articles about the implications of AVs.

| Implication | Keyword |
|---|---|
| Background | History, background, evolution, adoption |
| Safety | Safety, accident(s), collision(s), crash(es), risk(s), concerns |
| Liability/insurance | Liability, insurance, responsible, ownership |
| Privacy | Privacy, data, data protection, personal data, connected vehicles, location, tracking, surveillance |
| Cyber security | Cybersecurity, hacker(s), hacking, attack(s), cyber attack(s) |
| Unemployment | Economy, economic, jobless, mass unemployment, displace, taxi, drivers |
| Congestion | Congestion, jam(s) |
| Environmental effects | Fuel, fuel economy, fuel efficiency, emission(s), carbon emission(s), energy, energy use, pollutant(s) |
| Ride-sharing | Shared, shared vehicle(s), ride-sharing, uber, grab, lyft |
| Land use | Land-use, parking, infrastructure |
| Costs | Cost(s), price, parking costs, tax(es), affordable |
| Government regulation | [country name] regulation(s), legislation(s), rule(s), bill |
| Level of automation | Level of automation |

## Background to AVs

Autonomous systems are characterised as systems capable of making decisions independently of human interference (Brodsky, 2016; Collingwood, 2017), but, unlike mere automation, they can make these decisions while facing uncertainty (Danks & London, 2017). Autonomous systems have been developed in different domains, including warfare, personal care (Arkin, 2013; Pineau, Montemerlo, Pollack, Roy, & Thrun, 2003; Stahl & Coeckelbergh, 2016; Sukman, 2015) and transport. AVs rely on artificial intelligence (AI), sensors and big data to analyse information, adapt to changing circumstances and handle complex situations as a substitute for human judgement, as the latter would no longer be needed for conventional vehicle operations such as lane-changing, parking, collision avoidance and braking (Long, Hanford, Janrathitikarn, Sinsley, & Miller, 2007; West, 2016).[1] This perceived superiority to human drivers is attributed to high-performance computing that allows AVs to process, learn from and adjust their guidance systems according to changes in external conditions at much faster rates than the typical human driver, and it is supplemented with vehicle-to-vehicle (V2 V) and vehicle-to-infrastructure (V2I) communication, allowing AVs to learn from other vehicles (West, 2016).

Companies are racing to secure their share in the emerging AV market by investing in software development, teaming up with leading university research centres and implementing testing on roads. Most governments recognise the need to adapt to these rapid technological advancements but face challenges balancing the strategic desirability of AVs and the issues accompanying this technology. AVs entail enormous social and economic benefits. Countries committed to developing AVs desire greater mobility for the elderly and handicapped, as well as improved safety, and competitiveness in the automotive industry, including the UK, US, China, and Japan (Nikkei, 2017; West, 2016). China aims to lead the world in electric vehicles and AVs by 2030 (Dunne, 2016). Using AVs can boost productivity in countries facing labour shortages in the transport sector, such as Singapore (Lim, 2017) and Japan (Bloomberg, 2016b). AVs can also help meet other national objectives such as improving the fuel economy (Dunne, 2016) and reducing congestion and pollution (Hanai, 2018). However, AVs entail various risks. In this article, we focus on the governance of such risks.[2] More specifically, we examine the governance of technological risks as it broadly defines the unintended consequences arising from the technology.[3]

AVs are classified into different categories based on their features. The Society of Automotive Engineers (SAE) categorises AVs based on five levels of automation. At level 1 (assisted automation) and level 2 (partial automation), the dynamic driving tasks such as its operational and tactical aspects are performed by the human (SAE, 2014). From levels 3 to 5, all the dynamic driving tasks are performed by the automated driving system. At level 3 (conditional automation), the human driver is expected to control the vehicle occasionally. A vehicle is classified as fully autonomous at levels 4 (high automation) and 5 (full automation), but only at level 5 is the vehicle expected to drive itself under all environmental conditions (Milakis, van Arem, & van Wee, 2017). This definition is adopted by various national and international bodies, such as Australia's National Transport Commission (NTC) (Hillier et al., 2015; Sun et al., 2016), the UK's Department for Transport (DfT) (Clark et al., 2016), the US National Highway Traffic Safety Administration ((NHTSA), 2017), the Government of Ontario, Canada (Ticoll, 2015) and the European Road Transport Research Advisory Council (ERTRAC, 2017). This study focuses on AVs at SAE Levels 4-5, as they represent a greater fundamental shift in society.

## AV risks and governance strategies

Innovative technologies such as AVs create risks and unintended consequences that may decrease society's acceptance of them, which include environmental risk, market risk, social risk, organisational risk, political risk, financial risk, technological risk, and turbulence risks (Li, Taeihagh, & de Jong, 2018). This article focuses on technological risks, described as potentially negative social, physical, and economic consequences related to citizens' concerns in the adoption of novel technologies (Renn & Benighaus, 2013). Five types of technological risk are associated with AVs: safety, liability, privacy, cybersecurity, and industry influence.[4]

To ensure that society reaps the maximum gains from the emerging AV market, it is paramount for governments to introduce new measures and regulations to manage the risks associated with AVs. In this section, we explore the types of strategies adopted by various governments to govern the technological risks brought about by AVs. We employ a framework for identifying governance strategies for addressing these risks, and categorise them as *no-response*, *prevention-oriented*, *control-oriented*, *toleration-oriented,* or *adaptation-oriented strategies* based on the work of Li et al. (2018) and Li, Taeihagh, de Jong, and Klinke (forthcoming) (Table 2).

### *Safety*

At least 90% of vehicle accidents are estimated to be the result of human error (NHTSA, 2015; Smith, 2013; Sun et al., 2016). Adopting AVs can potentially reduce or eliminate the largest cause of car accidents while also outperforming human drivers in perception,

**Table 2.** Types of governing strategies and AV-related examples (adapted for AVs based on Li, Taeihagh & de Jong (2018) and Li et al. (forthcoming)).

| Strategy | Definition and AV examples |
| --- | --- |
| *No-response* | Policy-makers do not take any specific actions to address risks and may delay decisions due to their uncertain nature. In this scenario, policy-makers may not have any back-up plans or robust institutional frameworks to address impending threats. An example of this strategy in response to AV safety risks is when the government has neither established nor indicated its intentions to establish safety standards for AV manufacturers to follow during the testing of AVs. Another example is the US federal government not establishing any nation-wide rules regarding the allocation of liability and motor vehicle insurance. This strategy corresponds to fragile strategy (Duit & Galaz, 2008). No-response might also imply that policy-makers are ignorant about the potential negative consequences of risks. |
| *Prevention-oriented* | The main aim of this strategy is to avoid risks by taking preventive action. Prohibiting the adoption of innovative technologies is one such display of risk avoidance, as it seeks to prevent the existence of risk. One example is to temporarily prohibit or restrict AV testing on certain routes if a safety concern is identified (PennDOT, 2016). This strategy corresponds to risk minimisation strategy (Brown & Osborne, 2013) and is suitable to address risks of a more predictable nature, but is ineffective when risks are unexpected (Wildavsky, 1991). |
| *Control-oriented* | Policy-makers allow for the existence of risks, but take steps to control them by implementing formal policies and regulations (Osipova & Eriksson, 2013). Traditional methods of risk assessment are adopted to predict and regulate risks. One example of a control-oriented strategy is the Singapore government's response to AV safety risks. In 2017, amendments were made to the Road Traffic Act which now requires AV testers to pass safety assessments and developers to have robust accident mitigation plans before testing on roads (Road Traffic (Amendment) Bill, 2017). |
| *Toleration-oriented* | Policy-makers take action to ensure that the system or organisation's performance is robust to risks in a wide range of situations. One example of this strategy in response to AV safety risks is when the government introduces new legislation that requires all AV manufacturers to develop a comprehensive list of contingency plans that outline and justify the AV's responses to a diverse range of accident scenarios. Another example is the UK government's Vehicle Technology and Aviation Bill (HC Bill 143, 2017) that lays out a comprehensive list clarifying the liability of insurers and AV owners in the event of an accident and under a wide range of circumstances. This strategy corresponds to robustness and resistance strategies proposed by Nair and Howlett (2016) and Walker, Lempert, and Kwakkel (2013) respectively. Policy-makers also make forward-looking plans to mitigate potential consequences, such as by developing alternative solutions. |
| *Adaptation-oriented* | This strategy aims to improve the adaptive capability of the system or organisation. It emphasises on embracing uncertainty and improving its performance in response to shocks. Features of this strategy also include aspects of "forward-looking planning, joint responsibility", and "co-deciding" (Li et al., 2018)). This strategy corresponds to adaptive resilience and resilience strategies proposed by Nair and Howlett (2016), and Walker et al. (2013). For instance, Australia's National Transport Commission is seeking feedback from various stakeholders to decide on one of four options to regulate AV safety. Here, policymakers view risk as an opportunity to change the system for the better, rather than as a threat that should be ignored, suppressed, controlled, or tolerated. |

decision-making and execution. However, AVs introduce new safety issues. Collingwood (2017) and Litman (2017) highlight that vehicle occupants may reduce seatbelt use and pedestrians may become less cautious due to feeling safer. Also, the elimination of human error does not imply the elimination of machine error. As the technology grows in complexity, so does the probability of technical errors compromising vehicle safety. The fatal crash of Tesla's autopilot in 2016 reveals the uncertainty of machine perception (Banks, Plant, & Stanton, 2018) and highlights the technology's inability to avoid accidents in certain scenarios. Concerns also arise regarding how AVs should be programmed by "crash algorithms" to respond during unavoidable accidents (Coca-Vila, 2018; de Sio, 2017; Nyholm & Smids, 2016). Due to the "lack of blame", the damage caused by AVs in accidents cannot be assessed subjectively, which necessitates rules to regulate AVs' reactions to moral dilemmas (Coca-Vila, 2018). However, it is unclear how to arrive at these rules. Algorithms may be programmed to prioritise the safety of the AVs' occupants

"over anything else", which ensures the economic viability of developing AVs, but using the individual self-interest of AV occupants as a basis to justify the harm inflicted on others undermines the functions of law itself (Coca-Vila, 2018). In contrast, algorithms may be programmed to achieve the most socially beneficial decision based on a range of factors, but how to arrive at these factors is still unclear (Coca-Vila, 2018). Also, regulators have yet to agree on an acceptable level of safety or define legitimate methods of determining the safety of AVs (Kalra, 2017). AVs' performance could improve over time with real-world driving experience, but this is only possible if the public accepts the technology (Bansal, Kockelman, & Singh, 2016; Kalra, 2017).

In the US, the federal government traditionally sets the "national safety standards", and the state governments issue licences and regulate drivers' behaviour (Halsey, 2018). NHTSA outlined a Vehicle Performance Guidance for all entities involved in "manufacturing, designing, supplying, testing, selling, operating, or deploying" AVs in the US (NHTSA, 2017). While NHTSA has intentions to enforce these recommendations in future, now it requests these entities to provide a Voluntary Safety Assessment that outlines the compliance to the guidance, which includes specifications on systems safety such as describing safety strategies and design redundancies for addressing AV malfunctions (NHTSA, 2017). The responsibilities of the federal and state governments were clarified in the "Self Drive Act" in late 2017, which establishes NHTSA as the "preeminent regulating body" (Stone, 2018) and allows states to enforce new standards on AVs only if they are "identical" to what is prescribed by federal law (H.R.3388, 2017). It seems with AVs, the legal competence of the federal government will grow while that of state governments' shrinks (Halsey, 2018), as the role of the latter in regulating driver behaviour becomes more redundant. The federal government is not interested in imposing strict regulations on AVs, as, in the words of the Transportation Secretary, they are "not in the business … to pick the best technology" and prefer a market-oriented approach (Halsey, 2018).

Similarly, the UK's DfT published an AV testing code of practice for manufacturers to ensure AV safety in various situations throughout their service life (DfT, 2015), which also has no legal status. It encourages and allows testing on any public road in the UK without requiring the approval of authorities or a surety bond (CCAV, 2016). However, frameworks on how risk can be minimised while engaging in public testing have not yet been established. This laid-back approach stems from the plans to create a national "cluster of excellence" in AV testing as part of its Industrial Strategy to grow human capital, attract foreign investment and develop "high-skill, well-paying jobs" to enhance the economy and achieve greener economic growth, greater mobility, and meeting the needs of an ageing society (DBEIS, 2017a, 2017b). Both the US and UK are careful not to impose regulations that are too stringent, or to have an excessively lenient stance on AV safety, to provide sufficient room for innovation (CCAV, 2016; Kang, 2016). Their attempts to establish and align expectations regarding safety standards without imposing overly restrictive barriers to innovation represent a light control-oriented strategy.

Likewise, Australia's NTC has published non-mandatory guidelines for safe AV testing that also constitute a light control-oriented strategy (NTC, 2017b). In 2016, the Transport and Infrastructure Council approved of the NTC's suggestion to create a national safety assurance system to assess the level of safety of AVs (NTC, 2016; NTC, 2017c). Emphasis is placed on controlling access to AVs, and it supports the commercial deployment of AVs as a long-term goal, while no regulations have yet been established to approve deployment, and it will still be considered case by case (NTC, 2016). The NTC has developed four regulatory options to regulate safety, on which it is seeking feedback from various stakeholders (NTC, 2017c). This step represents an attempt at consensus-building and public participation among various actors and may thus reflect a move towards an adaptation-oriented strategy.

China's government also adopts a light control-oriented strategy to address safety risks while taking some preventive measures to avoid exposing AVs to realistic road conditions. Human drivers are required to be in the vehicle with their hands kept on the steering wheel, and AVs cannot be tested under actual road conditions until the government devises a framework for granting road test exemptions (KPMG, 2018; West, 2016). While the government has developed draft rules to regulate AV testing on public roads, AV testing has remained slow as existing laws have yet to be revised (The Straits Times, 2018). In 2016, the National Technical Committee of Auto Standardisation started reviewing China's vehicle standards and regulations to identify the appropriate regulatory adjustments. In 2017, the China-New Car Assessment Programme was initiated to ensure that safety measures are well incorporated into the assessment system, and research has begun on industry policy and stakeholder engagement of AVs to assist authorities (ERTRAC, 2017). AVs have been identified as a key sector in the government's plans in becoming a leader in artificial intelligence by 2025 and to compete with the US' core AI industries. Thus, China seeks to create a "friendly policy environment" for accelerating AV development (Cadell & Jourdan, 2017; Dai, 2018).

In Europe, AV testing is legally permitted, but the EU is stricter relative to the US due to cultural differences, as Europe emphasises more on protecting citizens from technological risks while the US focuses on the "race for innovation and progress" (Nicola, Behrmann, & Mawad, 2018). AV testing in the US is allowed on public roads without any mandatory standards to follow, while in Europe AV testing is typically "confined to private streets" and "pre-defined routes" or "restricted to very low speeds" (Nicola et al., 2018). Amendments to the 1968 Vienna Convention on Road Traffic took effect in 2016 to legalise the use of automated driving technologies, which the German government has incorporated into its national law in December 2016. The amended 1968 Vienna convention, however, still requires every vehicle to have a driver who should always be ready to take control of the AVs. The European Parliamentary Research Service (EPRS) highlights that this is incompatible with most highly or fully automated systems, which may not require a driver. Thus, the EPRS recommends further amending the convention (Pillath, 2016). The German government has started experimenting with safety standards through its project PEGASUS (FMEAE, 2017). At both the EU and national level, European governments are still evaluating the implications of AVs before establishing permanent regulations. The aim is to develop a unified strategy to regulate AVs, marked by the Declaration of Amsterdam in 2016, agreeing to meet twice a year to share best practices, monitor progress and collaborate on all levels of regulation (ERTRAC, 2017).

Singapore and Japan have begun amending their laws to regulate safety in AV testing. The Singapore Road Traffic Act (RTA) was amended in February 2017, demonstrating a control-oriented strategy. The law now recognises that a motor vehicle need not have a human driver (RTAB, 2017) and the Minister for Transport can create new rules on AV trials, set standards for AV designs, and acquire the data from AV trials. A five-year regulatory sandbox was created to ensure that innovation is not stifled and the government intends to enact further legislation in the future. Meanwhile, AVs must pass safety assessments, robust plans for accident mitigation must be developed before road testing, and the default requirement for a human driver can be waived once the AV demonstrates sufficient competency to the Land Transport Authority (LTA). After displaying higher competencies, AVs can trial on increasingly complex roads (CNA, 2017). Similarly, Japan has drafted rules for AV testing in early 2017 that require a human driver with a driver's licence in the vehicle, police approval, clear labelling on AV test vehicles and testers to always be prepared to apply brakes (Kyodo, 2017). Furthermore, police officers will "ride the test vehicles" to ensure its proper functioning (Jiji, 2017). The emphasis on human control of the AV demonstrates a prevention-oriented strategy, as the Japanese government is actively using human oversight to avoid the risk of accidents resulting from technical faults. In South Korea, a Smart Car Council has been established to coordinate actions across ministries (West, 2016).

### Liability

In most conventional car accidents, the driver retains some control over the vehicle and thus assumes primary liability for the vehicle's fate; however, persons in an AV are no longer in control (Collingwood, 2017; Douma & Palodichuk, 2012). Part or all of the responsibility will shift onto the AV as accidents become more of an issue of product safety or efficacy; thus, third parties involved in the design of safety systems in AVs will face greater vulnerability to lawsuits involving product liability (Marchant & Lindor, 2012; Pinsent Masons, 2016). It is unclear how liability will be apportioned between the AV's autonomous system and the human driver. Will the human bear part of the responsibility of a crash if there is a manual override function they failed to use (Collingwood, 2017)? At the expense of privacy, black box data (event data recorders (EDRs)) can be utilised for determining liability more accurately (Dhar, 2016). Moreover, no clear legal framework exists that outlines how liability is apportioned between third parties responsible for designing AV systems – the manufacturer, supplier, software provider or the software operator –making the identification and separation of the various components that caused the malfunction difficult (Collingwood, 2017; Pinsent Masons, 2016).

Manufacturers are increasingly vulnerable to reputational risks imposed by accidents associated with failures in design and manufacturing (Hevelke & Nida-Rümelin, 2015; Tien, 2017). The current legal frameworks also do not define the practical and moral responsibilities of software programmers in designing "crash algorithms" that determine life or death decisions, raising numerous concerns over AVs' implications on public ethics (Fleetwood, 2017; Pinsent Masons, 2016). Governments have yet to address whether algorithms' decision-making criteria during accidents should be standardised. For instance, should decisions be prioritised by the likelihood, severity, and quality of life effects of the type of injury, or by the number of people injured (Fleetwood, 2017)? No government save the UK has yet amended their legal framework to incorporate these new complexities into the liability of drivers, manufacturers, software designers and other third parties (Duffy & Hopkins, 2013; HC 143, 2017).

The assignment of liability and the corresponding effects on insurance costs are currently unknown (Abdullah, 2016b). Injured third parties may resort to suing the manufacturer or software provider if responsibility belongs to the autonomous system. In the long run, high liability risks may weaken the incentive for manufacturers to innovate, slowing down further safety improvements for AV users (Gurney, 2013; Hevelke & Nida-Rümelin, 2015).

In the US, the federal government delegates most of the responsibility in determining liability rules to state governments (NHTSA, 2017). Currently, the Department of Transportation has not displayed any response to establishing nation-wide rules for liability and insurance in the short run. NHTSA urges states to consider liability allocation, to determine who must carry motor vehicle insurance and to consider rules allocating tort liability. So far, most states have taken the first step towards a control-oriented strategy to address liability risks by revising the definitions of AVs (NHTSA, 2017).

The only country that has adopted a toleration-oriented strategy to address liability and insurance risks is the UK at the moment, and other countries have adopted either no-response or control-oriented strategies. At the end of 2016, the Centre for Connected & Autonomous Vehicles (CCAV, 2016) highlighted the legal gaps involving liability and insurance and proposed regulatory changes to the DfT. In response, the Bill HC 143 (2017) was passed. The bill lays out a comprehensive list clarifying the liability of insurers and AV owners if an accident occurs and under a wide range of circumstances. Insurers are automatically liable for death or damages due to accidents caused by insured AVs (HC 143, 2017). An insurer's liability can, however, be limited in situations where the owner is deemed at fault. The bill thus resolves ambiguity regarding the apportioning of liability between insurers and the insured victims involved in AV accidents. Specifically, the bill ensures that liability for accidents involving AVs remains under the existing motor vehicle insurance scheme, providing accident victims faster access to compensation (CCAV, 2016; DfT, 2017b). Manufacturers are also protected under the Consumer Protection Act if they demonstrate that the vehicle was not defective at the time it was supplied, and that the defect was only detected later due to scientific advancements (Coates, 2017).

Governments in Singapore and Australia have acknowledged the need to update liability laws. The Singapore government amended the RTA in 2017 to exempt AVs, its operators and those involved in AV trials from existing provisions of the RTA, which hold a human driver responsible for the use of vehicles on public roads (CNA, 2017). There is clear acknowledgement that the vehicle is now in the control of the AV system, and that AVs confront the notion of human responsibility at the core of current road and criminal laws in Singapore (MOT, 2017). In Australia, the government plans to follow a stated timeframe for amending liability and insurance laws. The NTC plans to develop guidelines clarifying the different definitions of control for AVs by November 2017. After this, it is committed to review current driving laws, establish specific legal obligations for AV driving entities, and, if necessary, amend compulsory injury systems to identify potential barriers to eligibility of occupants and accident victims by 2018 (NTC, 2017a). Overall, government efforts both in Singapore and Australia reflect a gradual approach towards regulatory reform and, thus, a movement towards a light control-oriented strategy to manage liability and insurance risks.

Currently, governments in China and South Korea have not indicated their regulatory stance towards liability and insurance risks, representing a no-response strategy. Notably, Baidu Inc. and automaker Zhejiang Geely Holding Co. have urged the Chinese government to speed up the drafting of regulations for AV testing (Bloomberg, 2016a).

The government in South Korea has mentioned that the lack of international standards is hindering the creation of domestic rules for AVs, as South Korea is a major importer and exporter of cars, requiring manufacturers to incorporate international standards for AVs (Ramirez, 2017).

The EU has not amended its legal framework to incorporate AV-related liability and insurance risks but is exploring solutions to liability issues. The European Commission (EC) launched GEAR 2030 in 2016 to explore solutions to AV-related issues, and in February 2017 the group made recommendations for using EDRs. In May 2016 European Parliament Members recommended that the EC should create a mandatory insurance scheme and an accompanying fund to safeguard full compensation for victims of AV accidents and a legal status should be created for all robots to determine liability in accidents (EP, 2017; EPCLA, 2016).

Like the UK, the government of Germany has enacted permanent legislation in June 2017 to address AV-related liability risks. According to the law, AVs must install a black box to record the entire journey to determine liability during collisions (JDSUPRA, 2017; Wacket, Escritt, & Davis, 2017). The law also doubles the maximum liability limits imposed by the existing RTA and attempts to apportion liability between the manufacturer and the driver: the former is made responsible for accidents where the AV system is in charge, and a system failure is the main culprit (Wacket et al., 2017. However, the law lacks clarification on what is considered an "adequate time reserve" that drivers are permitted to have before taking control when necessary and on what grounds third parties own the data collected in the black box (JDSUPRA, 2017). Germany's new Ethics Commission has also published the world's first ethical guidelines for AVs. The guidelines recommend that there must always be clarity regarding who is considered the driver, which must be documented for determining liability. Moreover, it states that it is unethical for algorithms in the AV system to use an individual's data (such as their age or gender) as criteria for decision-making during unavoidable accident scenarios (FMTDI, 2017a). Although the guidelines are not mandatory, it is a first step towards resolving the ethical issues surrounding AVs. There has yet, however, to be an open discussion regarding the responsibility of persons designing such algorithms.

Japan's strategy towards liability and insurance risks can be classified as light control-oriented. The National Police Agency makes recommendations on actions to avoid liability risks but has not made them mandatory. For instance, it urges companies to install black boxes on AVs that are tested to help ascertain the causes of accidents and take preventive measures (Nikkei, 2018). Also, testers of AVs are required to submit documents detailing the structures of vehicles and accident mitigation plans to the authorities. The operators or monitors of AVs through remote systems must have a driver's licence and bear responsibility for operational mistakes (Jiji, 2017; Japan Bullet, 2017). Manufacturers will be liable for defects in the system, but this does not include the software designer or other third parties involved in the initial design of the vehicle (Japan Bullet 2017).

## *Privacy*

AVs are reliant on sensors, high definition maps and other instruments, from which information is collected and optimised to ensure the vehicle's safe operation (West, 2016; Dhar, 2016). However, concerns arise regarding who controls this information, and how it is used (Anderson et al., 2014; Boeglin, 2015). Multiple issues regarding informational privacy remain unclear: the exact reasons why information is being collected, the types of information being collected, accessibility to the information and the permissible duration of information storage have not been clarified (Glancy, 2012). V2V and V2I communications allow information to be transmitted between AVs for safety reasons, but they also expose the vehicle's movements and geographical location to external networks, from which people can access to locate an AV user (Glancy, 2012). Schoonmaker (2016) highlights the inadequacies of protecting location-based data based on customer consent, as customers accept the terms and conditions without fully understanding them. Another issue is the use of EDRs for ascertaining the exact causes of accidents, as this data may be sold to third parties such as insurance companies and used against drivers (Dhar, 2016; Pinsent Masons, 2016; Schoonmaker, 2016).

Other cited risks to informational privacy are the possibility of using this information to harass AV users through marketing and advertising, to steal users' identity, profile users and predict their actions, concentrating information and power over large numbers of individuals (Glancy, 2012). While it is possible to anonymise the information taken, this can be reversed through deanonymisation.[5] Deanonymisation algorithms can re-identify anonymised microdata with high probability, demonstrating that anonymisation is insufficient for data privacy (Gambs, Killijian, & del Prado Cortez, 2014; Narayanan & Shmatikov, 2008). This is a serious problem for location-based data, as human traces are unique, enabling an adversary to trace movements even with limited side information (Gambs et al., 2014; Gillespie, 2016). Also, access to the interconnected[6] AVs' wireless network enables public and private agencies to conduct remote surveillance of AV users, which can undermine individual autonomy through psychological manipulation and intimidation (Glancy, 2012). Another emerging issue is the use of video surveillance in AVs that are used as a transportation service, such as autonomous taxis. As users do not own these AVs, it is unclear whether the vehicle is considered a "public space" where surveillance can be considered acceptable (Schoonmaker, 2016).

The governments in the US and South Korea have enacted new legislation on data privacy that applies to all vehicles (including AVs and conventional vehicles). In the US, the new SPY Car Act gives NHTSA the authority to protect the use of (and access to) driving data in all vehicles manufactured for sale in the US (SCA, 2017). All vehicles must provide owners or lessees the ability to stop the data collection, except for data essential for safety and post-incident investigations, and manufacturers are prohibited from using the collected data for marketing or advertising without consent from the owners or lessees. Similarly, effective on February 2016 the South Korean government amended its Vehicle Management Act which establishes conditions for the issuance of temporary licences to test AVs and sets requirements on data collection for all vehicles. Any individual must obtain approval from the Minister of Land, Infrastructure and Transport (LIT) before using collected data. The Act does not, however, specify the extent of information sharing in different conditions. It mentions that approval will be granted in a way that does not violate the privacy of vehicle owners, and that the standards for approval will be determined by the Minister of LIT (MVMA, 2017).

The EU has taken steps to manage privacy and cybersecurity risks applicable to all data in the region, demonstrating a control-oriented strategy. The European Parliament's Intelligent Transport Systems (ITS) Action Plan in 2009 (EP, 2009) emphasised the need to protect personal privacy from the early stages of designing ITS, and the EC released a study in 2012 assessing possible methods to ensure data protection in ITS (Pillath, 2016). These efforts were consolidated through the Declaration of Amsterdam (MIE, 2017). The Data Protection Directive 95/46/EC of 1995 was then updated through the EU General Data Protection Regulation (EU GDPR), which was ratified in 2016 and will become effective in May 2018.[7] The regulations will apply to all companies processing data from subjects residing in the EU, regardless of the location of the company, extending control of data beyond geographical borders (EU, 2016). The regulations also strengthened conditions for consent and increased penalties to a maximum fine of 4% of companies' global revenue and protects the right to be forgotten and the "right to explanation", which allows citizens to review particular algorithmic decisions (Metz, 2016). The EU has already fined Google on several occasions, demonstrating its commitment to privacy (Eben, 2018; West, 2016). However, stringent application of these rules may impede AV developments, for instance, high definition mapping requires geo-coded data to improve AVs' navigational abilities. Excessive regulation of data usage may also disadvantage European manufacturers, and it may be difficult to enforce the GDPR on non-European manufacturers (Pinsent Masons, 2016).

China and Japan have both also taken legislative action to control privacy and cybersecurity risks applying to all personal data, demonstrating a control-oriented strategy. For instance, Japan has amended its Privacy Protection Law in 2017 (The Japan Times, 2017). China, too, has enacted a new Cybersecurity Law requiring the anonymisation of all forms of personal information. It emphasises customer consent and requires network operators to be transparent regarding the purpose, method, and scope of data collection and use (KPMG, 2017). Overall, the law establishes many controls on the collection, use and sharing of personal data but the law does not, however, include AV-specific provisions.

The Singapore government adopts a control-oriented strategy to address privacy risks in general and specifically between public sector agencies. The government is in the process of amending the Personal Data Protection Act (PDPA). A public consultation was issued in July 2017 that proposes amendments to the PDPA, such as increasing the transparency regarding the collection and use of personal information and providing individuals with the option to terminate their consent of these data collection activities (PDPC, 2018). Also, the government has enacted the Public Sector (Governance) bill that prohibits the unauthorised use and sharing of data between public sector agencies. The bill is designed to improve the delivery of public services in Singapore, particularly in the aspects of efficiency and "programme management" (PSGB, 2017).

Germany and Australia have not amended existing legislation to address AV-specific privacy risks. Germany's new AV bill does not include provisions for data privacy but addresses safety and liability risks. The German government has, however, indicated its intention to incorporate privacy concerns when the bill is revised in two years (Wacket et al., 2017). Australia's NTC has also released privacy recommendations, such as adopting a "privacy by design" approach and refraining from generating personal information "wherever possible"; however, this last phrase may suggest that these recommendations are rhetorical overtures (Daly, 2017). Thus, these principles may represent a formal commitment to risk control rather than specifically outlining steps to control AV privacy risks. The NTC also recommended that the upcoming national safety assurance system incorporate elements of privacy protection at the highest possible level (NTC, 2017a). More recently, the House of Representatives Standing Committee on Industry, Innovation, Science and Resources (SCIISR) encouraged public participation in an inquiry into the social implications of AVs and recommended further investigating the data rights of consumers, insurers, government agencies, and manufacturers (NTC, 2017a), adopting an adaptation-oriented strategy by engaging with the public to build consensus in addressing privacy risks.

The UK's DfT, in collaboration with the Centre for the Protection of National Infrastructure (CPNI), created key principles for privacy and cybersecurity. The guidelines recommend that manufacturers follow ISO standards, such as the Privacy Architecture framework outlined by ISO 29101 (DfT, 2017a), demonstrating a light control-oriented strategy. The principles state that personal information must be "managed properly" concerning what is stored and transmitted, its usage, the data owner's control over these processes and ensuring AV users' ability to delete "sensitive data". However, what is considered "proper" management of personal information or "sensitive" data is not defined. These efforts indicate the government's awareness of AV-specific privacy risks and the non-binding nature of the guidelines supports the government's aspirations in becoming a world-leading hub for AV research and development and thus not taking actions that may impede achieving this aim (DfT, 2017a).

In Germany, 13 voluntary recommendations for AVs have been released, and notably, it is recommended that specific rules clarify the data that businesses can process without the "explicit consent" from the AV users (FMTDI, 2017b). Similar to the EU's GDPR, these recommendations apply to all data and emphasise on complete transparency and drivers' full authority over the use of personal data collected from the AV. Germany's current data protection laws are strict regarding the definition of personal data as applied to information with the slightest link to an individual and it is likely that most connected AV data will be considered as personal data unless data-generating items have been designed to anonymise data (Pinsent Masons, 2016).

## *Cybersecurity*

Cybersecurity threats to conventional vehicles with automated features already exist. In their survey of 5000 respondents across 109 countries, Kyriakidis, Happee, and de Winter (2015) found that people were most concerned about software hacking and misuse of vehicles with all levels of automation. Hackers could take control of the vehicle through wireless networks (such as Bluetooth, keyless entry systems, cellular or other connections) as the car connects with the environment (Lee, 2017). With its ability to store and transmit transaction and lifestyle data, AVs are attractive targets for hackers as such information can be sold for a financial gain, or these systems can be used to inflict physical harm by extremists or used for illegal purposes by drug traffickers (König & Neumayr, 2017; Lee, 2017). For instance, Miller and Valasek demonstrated that malicious attacks on AVs are a near-term possibility in 2013, as they hacked a Chrysler-Jeep through its internet connection and took control of its engines and brakes (Schellekens, 2016).

Various studies have analysed the possible cybersecurity threats to AVs, as computers possess greater control over the movements of an AV, AVs are more vulnerable to hacking than conventional vehicles, and the driver is less able to intervene during an attack (Hern, 2016; Lee, 2017). Without sufficient security, V2V and V2I communication channels can be hacked, which can lead to serious accidents (Dominic, Chhawri, Eustice, Ma, & Weimerskirch, 2016; Pinsent Mason, 2016). Injection of fake messages and spoofing of global navigation satellite systems (GNSS) are some of the major threats that AVs will face, as GNSS data can be manipulated to undermine the AVs' safety critical functions (Bagloee, Tavana, Asadi, & Oliver, 2016). Other threats include the use of sensor manipulation to disorient the AV's systems, bright lights to blind cameras and ultrasound or radar interference to blind an AV from incoming obstacles (Page & Krayem, 2017; West, 2016). While systems may be installed to detect such malfunctions, these require software updates as well as changing existing standardised security architectures (Bagloee et al., 2016).

Most governments have developed non-mandatory guidelines on cybersecurity best practices and researched to explore the implications of AVs on cybersecurity. Governments in the US, China, EU, and Singapore have adopted a control-oriented strategy and have introduced or enacted new legislations to address cybersecurity risks.

In the US, NHTSA's voluntary guidelines recommend that manufacturers and software companies design AV systems according to existing international standards, such as those published by the National Institute for Standards and Technology, NHTSA, SAE and the Alliance of Automobile Manufacturers and others (NHTSA, 2017). A new electronics systems safety research department has been set up to evaluate and monitor potential cyber vulnerabilities and an internal agency working group, the Electronics Council, has also been set up to enhance collaboration regarding electronics and cybersecurity research (NHTSA, 2018). These changes represent attempts to gain more awareness and raise awareness of cybersecurity risks to automakers and software companies. The SPY Car Act was also introduced to enhance controls on cybersecurity and privacy to all vehicles (SCA, 2017).[8] According to this law, critical and noncritical software systems in every vehicle must be separated, and all vehicles will be evaluated using best practices. It introduces specifications to ensure the security of collected information in vehicle electronic systems while the data is on the vehicle, in transit from the vehicle to a different location or in any offboard storage. It also requires vehicles to be able to instantaneously detect, stop and report attempts to capture driving data or take control of the vehicle and requires the AV to display the extent to which the AV protects the privacy and cybersecurity of the consumers.

Cybersecurity is not a new concern in the EU. It has taken incremental steps to control cybersecurity risks over the last few years, although they are not AV-specific. The EU Cybersecurity strategy was introduced in 2013, followed by the Directive on the security of network and information systems in 2016 (EC, 2017). The latter was the first EU-wide legislation on cybersecurity. Further efforts have been taken by various EU organisations to raise awareness and provide recommendations on how to address cybersecurity issues. In 2016, the EU's independent advisory body on data protection and privacy, the Data Protection Working Party, published its views to raise awareness about developments in the IoT and its associated security issues (Pillath, 2016).

Like Europe's GDPR, China's latest Cybersecurity Law represents a control-oriented strategy. Key provisions of the law are personal information protection, critical information infrastructure protection, responsibilities of network operators to ensure security, preservation of sensitive information within China, certification of security products and penalties for violations (KPMG, 2017). One example of network operators' responsibilities includes the requirement for critical information infrastructure operators to store personal data within China and for companies to gain approval and pass national reviews before moving data overseas (He, 2018). Critical cyber equipment and special cybersecurity products can only be sold after receiving security certifications (KPMG, 2017). The government in Singapore has also amended existing legislation to control different aspects of cybersecurity risks. Singapore's Computer Misuse and Cybersecurity Act was amended in April 2017 to strengthen businesses' response to computer-related offences (Kwang, 2017). Other steps have been taken to raise awareness of cybersecurity, such as through local institutes of higher learning and forming partnerships between academia and the private sector. By doing so, the government aims to use this as an opportunity for Singapore to become a leading cybersecurity service provider, demonstrating an adaptation-oriented strategy; and there are plans to set up a national Defence Cyber Organisation (Srikanthan, 2017).

The UK government has not yet exerted legal control over cybersecurity risks in AVs but is taking steps to increase awareness and strengthen the resilience of AVs against such risks. It has implemented two cybersecurity strategies applying to all cyber systems in the UK. The National Cybersecurity Strategy 2016–2021 focuses on promoting further research into cybersecurity for all systems to produce successful products and services and strengthen UK's position as a world leader in cybersecurity by 2021 (Cabinet Office, 2016). A National Cybersecurity Centre (NCSC) was established in 2016 to analyse and detect cyber threats. The strategy also targets autonomous systems, which may receive funding for research in the upcoming Cyber Science and Technology Strategy (Cabinet Office, 2016). Besides, the strategy aims to stimulate growth in the cybersecurity sector and to enhance its citizens' responses to these threats, which represents an effort to enhance the country's adaptive capacity. The government's adaptation-oriented approach is also reflected in the DfT and the CPNI's key principles for privacy and cybersecurity, which recommends designing the AV system to be resilient to attacks and to produce appropriate responses when its defences or sensors fail (DfT, 2017a).

South Korea has amended its Vehicle Management Act, but it does not include provisions related to AV cybersecurity (MVMA, 2017). Australia and Germany have not amended legislation on cybersecurity but are exploring the security risks arising from AVs. The government in Germany has set up 5 working groups to research AV-related issues such as cybersecurity and data protection (ERTRAC, 2017). More recently, in Australia, it was recommended that the National Cybersecurity Strategy investigate AVs and associated systems to address potential vulnerabilities and recommends establishing a national taskforce to coordinate the introduction of AVs (SCIISR, 2017). The Japanese government appears to have adopted a no-response strategy; as it has neither amended its RTA nor provided recommendations on either general or AV-specific cybersecurity risks. The government has, however, displayed intentions to gain more awareness and revise laws on liability and cybersecurity issues (Nikkei, 2015).

### *Industry influence*

Literature suggests that technological advancements pose a threat to many existing low-skilled, manual jobs, as these are easily automated (Brynjolfsson & McAfee, 2011; Frey &

Osborne, 2017). Drivers and mechanics are especially at risk as their value-added is derived from the driving task and they tend to be older and less educated (Alonso Raposo et al., 2018). If the regulatory environment favours widespread adoption, AVs will have immense employment implications. Simulation studies suggest that taxi fleets could be reduced in size to 10% in Berlin, and to one third in Singapore if autonomous taxi services also replaced traditional public transport (Bischoff & Maciejewski, 2016; Spieser et al., 2014). In Singapore, where start-up nuTonomy launched driverless taxis for the first time in the world, nearly half of the privately-owned cars may be redundant in future (Liang & Durbin, 2016). Truck drivers and bus drivers are also at risk due to the massive cost savings from eliminating labour (Anderson et al., 2014; Clements & Kockelman, 2017; Frey & Osborne, 2017). It is estimated that the trucking and delivery industries will gain $100–$500 billion from AVs by 2025, most of which will come from eliminating drivers' wages; while shifting truck drivers to more technical roles, such as monitoring AV systems, will barely make up for the millions of jobs lost (Clements & Kockelman, 2017). Overall, the net economic effects of introducing AVs are estimated to be positive, but the redistribution of employment will negatively impact lower-skilled workers the most, as these displaced workers may spillover to other low-skilled occupations, creating downward pressure on their wages, which can exacerbate inequality (Alonso Raposo et al., 2018).

A few countries recognise the threat AVs pose to employment, although they have yet to formulate detailed strategies to address them. The US Transportation Secretary has voiced her concerns over the impact of AVs on employment (Reuters, 2017). In Australia, the SCIISR (2017) noted concerns about the negative implications of automation for professional drivers and acknowledged the impact of AVs on other sectors, such as the motor trades sector, insurers, repairers, and road enforcement officers. To minimise these potential negative effects, it urged transitioning the workforce as soon as possible. Singapore's government has conveyed its intention to retrain future displaced workers progressively through programmes helping them acquire new skills and enabling them to get higher value-added jobs (CNA, 2017). Much emphasis is placed on helping the Singaporean workforce to cope and adapt to inevitable disruption. Autonomous buses can fill up the shortage of bus drivers (CNA, 2017), and AVs can be used for street-cleaning purposes (Abdullah, 2016a); thus, the risk of disruption to employment in the public transportation sector is low relative to other countries without manpower constraints. These public statements signal the Singapore government's intentions of transforming AV-specific risks to employment into a beneficial opportunity for the nation's economy, demonstrating an adaptation-oriented strategy.

**Table 3.** Summary of the governing strategies adopted.

| Countries | Safety | Liability | Privacy | Cybersecurity |
|---|---|---|---|---|
| | | Risk Type | | |
| US | Light control | No response[a] | Control | Control |
| UK | Light control | Toleration | Light control | Adaptation |
| Australia | Light control, Adaptation | Light control | Light control, Adaptation | No response[b] |
| EU | Light control | No response[c] | Control | Light control |
| Germany | Control | Light control, Control | Light control, Control | No response[d] |
| China | Prevention, Light control | No response | Control | Control |
| Singapore | Control | Light control | Control | Control, Adaptation |
| Japan | Prevention | Light control | Control | No response |
| South Korea | No response[e] | No response | Light control | No response |

[a]No response is referring to the federal government's response, whereas most states in the US have adopted a control-oriented strategy towards AV liability risks.
[b]Recommended research and establishing a coordinating body.
[c]Conducted research and recommended a new insurance scheme.
[d]Working groups conducting research.
[e]Created a council to coordinate actions across ministries.

## Conclusion

This study aimed to obtain an overview of the governance strategies adopted so far in various countries in response to AV developments. As the basis of our analysis of government responses, we identified different technological risks associated with AVs and focused on five categories of risks: safety, liability, privacy, cybersecurity, and industry influence. Table 3 highlights the strategies adopted by different countries for addressing these AV risks.

Research shows that AV-related safety risks may arise from the less cautious behaviour of vehicle occupants and road users, system errors, and the lack of regulation of crash algorithms that determine life or death situations during inevitable accidents. Safety performance may improve over time if the public accepts mass deployment, which would allow AVs to gain more real-world driving experience. In response, most national governments have avoided using overly stringent measures to manage safety risks and have adopted light control-oriented strategies in the form of non-mandatory AV testing guidelines with the aim of promoting AV development. Given that AV development is at an early stage, councils or working groups have been created to explore the implications of the technology. Germany and Singapore have advanced to implement new regulations whereas China and Japan are currently developing regulation to regulate safety in AV testing. Australia has sought public consensus to address AV safety risks, demonstrating a move towards an adaptation-oriented strategy.

Lack of clarity regarding how liability is apportioned between AV occupants, AV manufacturers and other third parties along the supply chain may increase liability and reputational risks for manufacturers during accidents. To address liability and insurance risks, most governments either display no response or have adopted light-control oriented strategies in the form of voluntary guidelines and exploring possible options to address these risks before enacting legislation. The UK's new law resolves significant ambiguity regarding liability and insurance implications of AVs under various accident scenarios, reflecting the government's toleration-oriented approach. Germany enacted a similar law that provides less clarity relative to the UK regarding the responsibilities of drivers and data ownership permissions and thus, reflects a control-oriented strategy.

The literature also highlights the privacy risks that emerge alongside AVs. Data storage and transmission capabilities allow third parties to gain access to the personal information of customers and use it for advertising, user profiling, and location tracking. Responses to manage privacy risks range from enacting new data privacy laws, relying on existing data privacy laws to making recommendations on privacy principles. The EU and governments in most countries have developed new regulations to control the access to, use and sharing of personal data that are not specific to AVs, whose provisions vary regarding the scope and the extent of control given to consumers, among other aspects. An exception is the governments of Australia and the UK, who have made privacy recommendations. Countries that adopt light control-oriented strategies intend to regulate AV privacy risks in future, reflecting a dominant pattern towards control-oriented strategies. Australia's government has also pursued the less common strategy of building consensus with the public to address privacy risks.

AV Communication networks are vulnerable to malicious attacks that undermine cyber and physical security. Responses to manage cybersecurity issues vary considerably among the surveyed countries ranging from amending or introducing new non-AV specific legislation, creating working groups to explore these issues, funding cybersecurity-related research in the private sector and providing Cybersecurity principles to manufacturers. The release of Cybersecurity principles reflect the government's intentions to gradually shape AV developments alongside technological progression before making any hasty policy decisions. The US, China and Singapore have enacted cybersecurity laws that are not specific to AVs, Germany and Australia are still gaining awareness of AV cybersecurity risks, whereas the UK and Singapore display intentions to use cybersecurity risks as an opportunity to improve the nation's adaptive capacity. Overall, the strategies taken by most countries to address Cybersecurity risks encompass all systems in general rather than being specific to AVs.

Our research shows that AVs can also disrupt the public transportation and trucking industry, as AVs can displace workers from jobs that are easily automated. Most governments have not responded to these risks, but Singapore has begun programmes to retrain workers who might be negatively affected, while some governments have begun studying and regulating other risks such as the risks AVs pose to the environment, congestion, and government revenues.

## Notes

1. AVs are also referred to as driverless vehicles, as they are expected to operate safely without supervision and in all environments (Lin, 2016).
2. For a comprehensive study of the societal implications of AVs, see Milakis et al. (2017a).
3. For a full definition of technological risks, see the section entitled "Risks of AVs".
4. In this article, we focus on safety, liability, privacy, cybersecurity, industry influence risks associated with AVs. This is because while AVs potentially pose other risks such as to levels of congestion, the environment, land use, public infrastructure investment, government revenues or even organ shortage (Brodsky, 2016; Milakis et al., 2017; Bischoff & Maciejewski, 2016; Clements & Kockelman, 2017), these risks have not received enough attention from governments (see more in the next Section).
5. Deanonymisation, the process of using "background knowledge and cross-correlation with other databases", allows an unauthorised person to re-identify individual data records (Narayanan & Shmatikov, 2008).
6. Interconnected AVs are connected to one or more external communication networks, and the ability to send and receive external information keeps the vehicle up to date with the immediate roadway environment, allowing it to engage with other vehicles on the road and negotiate manoeuvres which are purported to be advantageous for road safety and efficiency (Piao & McDonald, 2008).
7. The EU GDPR was first introduced in 2012 (EU, 2016).
8. A similar Cyber AIR Act has been passed for aircraft (Bender, 2017).

## Acknowledgements

Araz Taeihagh is grateful for the support provided by the Lee Kuan Yew School of Public Policy, National University of Singapore through the Start-up Research Grant. Authors thank the three anonymous reviewers and the SI editor for their helpful comments and Eduardo Araral for his encouragements.

## Disclosure statement

## Funding

This work was supported by Lee Kuan Yew School of Public Policy, National University of Singapore [Start-up Research Grant].

## ORCID

*Araz Taeihagh*  http://orcid.org/0000-0002-4812-4745

## References

Abdullah, Z. (2016a). Driverless cleaning vehicles in the works for Singapore streets. *The Straits Times*. Retrieved from http://www.straitstimes.com/singapore/driverless-cleaning-vehicles-in-the-works-for-singapore-streets

Abdullah, Z. (2016b). Driverless vehicles could change laws, insurance policies. *The Straits Times*. Retrieved from http://www.straitstimes.com/singapore/transport/driverless-vehicles-could-change-laws-insurance-policies

Alonso Raposo, M., Grosso, M., Després, J., Fernández Macías, E., Galassi, C., Krasenbrink, A., … Thiel, C. (2018). An analysis of possible socio-economic effects of a Cooperative, Connected and Automated Mobility (CCAM) in Europe. *European Union*.

Anderson, J. M., Nidhi, K., Stanley, K. D., Sorensen, P., Samaras, C., & Oluwatola, O. A. (2014). *Autonomous vehicle technology: A guide for policymakers*. Santa Monica, CA: Rand Corporation.

Arkin, R. (2013). Lethal autonomous systems and the plight of the non-combatant. *Artificial Intelligence and Simulation of Behaviour Quarterly, 137*. Retrieved from http://www.aisb.org.uk/publications/aisbq/AISBQ137.pdf#page=4

Bagloee, S. A., Tavana, M., Asadi, M., & Oliver, T. (2016). Autonomous vehicles: Challenges, opportunities, and future implications for transportation policies. *Journal of Modern Transportation*, *24*(4), 284–303.

Banks, V. A., Plant, K. L., & Stanton, N. A. (2018). Driver error or designer error: Using the perceptual cycle model to explore the circumstances surrounding the fatal tesla crash on 7th May 2016. *Safety Science*, *108*, 278–285.

Bansal, P., Kockelman, K. M., & Singh, A. (2016). Assessing public opinions of and interest in new vehicle technologies: An Austin perspective. *Transportation Research Part C*, *67*, 1–14.

Bender, D. J. (2017). Senators reintroduce cybersecurity legislation for cars and planes. *The National Law Review*. Retrieved from https://www.natlawreview.com/article/senators-reintroduce-cybersecurity-legislation-cars-and-planes

Bischoff, J., & Maciejewski, M. (2016). Simulation of city-wide replacement of private cars with autonomous taxis in Berlin. *Procedia Computer Science*, *83*, 237–244.

Bloomberg. (2016a). China bans highway testing of autonomous cars pending regulation. *Bloomberg*. Retrieved from https://www.bloomberg.com/news/articles/2016-07-19/china-bans-highway-testing-of-autonomous-cars-pending-regulation

Bloomberg. (2016b). SoftBank's self-driving buses are coming soon to Japan's country roads. *The Japan Times*. Retrieved from http://www.japantimes.co.jp/news/2016/09/07/business/tech/softbanks-self-driving-buses-coming-soon-japans-country-roads/#.WVRVJ8ap3-Z

Boeglin, J. (2015). The costs of self-driving cars: Reconciling freedom and privacy with tort liability in autonomous vehicle regulation. *Yale Journal of Law & Technology*, *17*, article 4, 171.

Brodsky, J. S. (2016). Autonomous vehicle regulation: How an uncertain legal landscape may hit the brakes on self-driving cars. *Berkeley Technology Law Journal*, *31*(2), 851–878.

Brown, L., & Osborne, S. P. (2013). Risk and innovation. *Public Management Review*, *15*(2), 186–208.

Brynjolfsson, E., & McAfee, A. (2011). *Race against the machine*. Lexington, MA: Digital Frontier Press.

Cabinet Office, National security and intelligence, HM Treasury, and The Rt Hon Philip Hammond MP. (2016). National cyber security strategy 2016–2021. Retrieved from https://www.gov.uk/government/publications/national-cyber-security-strategy-2016-to-2021

Cadell, C., & Jourdan, A. (2017). China aims to become world leader in AI, challenges U.S. dominance. Reuters. Retrieved from https://www.reuters.com/article/us-china-ai/china-aims-to-become-world-leader-in-ai-challenges-u-s-dominance-idUSKBN1A5103?il=0

CCAV. (2016). Pathway to driverless cars: Proposals to support advanced driver assistance systems and automated vehicle technologies. *Crown*. https://www.gov.uk/government/consultations/advanced-driver-assistance-systems-and-automated-vehicle-technologies-supporting-their-use-in-the-uk

Clark, B., Parkhurst, G., & Ricci, M. (2016). Understanding the socioeconomic adoption scenarios for autonomous vehicles. Project report. The University of the West of England, Bristol. http://eprints.uwe.ac.uk/29134

Clements, L. M., & Kockelman, K. M. (2017). Economic effects of automated vehicles. *Presented at the 96th annual meeting of the transportation research board*, January 2017.

CNA. (2017). Regulations in place to ramp up driverless vehicle trials in Singapore. *Channel News Asia*. http://www.channelnewsasia.com/news/singapore/regulations-in-place-to-ramp-up-driverless-vehicle-trials-in-sin-7622038

Coates, C. (2017). UK bill to set a precedent for autonomous car legislation. https://www.dwf.law/news-events/legal-updates/2017/03/uk-bill-to-set-a-precedent-for-autonomous-car-legislation/

Coca-Vila, I. (2018). Self-driving cars in dilemmatic situations: An approach based on the theory of justification in criminal law. *Criminal Law and Philosophy*, *12*(1), 59–82.

Collingwood, L. (2017). Privacy implications and liability issues of autonomous vehicles. *Information & Communications Technology Law*, *26*(1), 32–45.

Dai, S. (2018). "China's Google" says government support and population size matters in global AI race. *South China Morning Post*. Retrieved from http://www.scmp.com/tech/enterprises/article/2127386/baidu-updates-driverless-car-software-wake-dispute-former

Daly, A. (2017). Privacy in automation: An appraisal of the emerging Australian approach. *Computer Law & Security Review*, *33*(6), 836–846.

Danks, D., & London, A. J. (2017). Regulating autonomous systems: Beyond standards. *Expert Opinion, Carnegie Mellon University*.

DBEIS. (2017a). Government sets out next steps in establishing the UK as global leader in connected and autonomous vehicles. Department for Business, Energy & Industrial Strategy, UK. Crown. Retrieved from https://www.gov.uk/government/news/government-sets-out-next-steps-in-establishing-the-uk-as-global-leader-in-connected-and-autonomous-vehicles

DBEIS. (2017b). Industrial strategy: The grand challenges. Department for Business, Energy & Industrial Strategy, UK. Crown. Retrieved from https://www.gov.uk/government/publications/industrial-strategy-the-grand-challenges/industrial-strategy-the-grand-challenges

de Sio, F. S. (2017). Killing by autonomous vehicles and the legal doctrine of necessity. *Ethical Theory and Moral Practice*, *20*(2), 411–429.

DfT. (2015). The pathway to driverless cars: A code of practice for testing. Department for Transport *Crown*.

DfT. (2017a). The key principles of vehicle cyber security for connected and automated vehicles. Retrieved from https://www.gov.uk/government/publications/principles-of-cyber-security-for-

connected-and-automated-vehicles/the-key-principles-of-vehicle-cyber-security-for-connected-and-automated-vehicles

DfT. (2017b). New measures set out autonomous vehicle insurance and electric vehicle infrastructure. *Department for Transport.* Retrieved from https://www.gov.uk/government/news/new-measures-set-out-autonomous-vehicle-insurance-and-electric-car-infrastructure

Dhar, V. (2016). Equity, safety, and privacy in the autonomous vehicle Era. *Computer*, *49*(11), 80–83.

Dominic, D., Chhawri, S., Eustice, R. M., Ma, D., & Weimerskirch, A. (2016). Risk Assessment for Cooperative Automated Driving. 2nd ACM Workshop on Cyber-Physical Systems Security and Privacy (pp. 47–58). Retrieved from https://pdfs.semanticscholar.org/61cc/e71b6ff9e83d6020f48d197ea5d85affc679.pdf

Douma, F., & Palodichuk, S. A. (2012). Criminal liability issues created by autonomous vehicles. *Santa Clara Law Review*, *52*, 1157–1169.

Duffy, S. H., & Hopkins, J. P. (2013). Sit, Stay, Drive: The Future of Autonomous Car Liability, SMU Science & Technology Law Review. 101. Retrieved from https://ssrn.com/abstract=2379697

Duit, A., & Galaz, V. (2008). Governance and complexity: Emerging issues for governance theory. *Governance*, *21*(3), 311–335.

Dunne, M. J. (2016). China aims to be No. 1 Globally in EVs, autonomous cars by 2030. Forbes. Retrieved from https://www.forbes.com/sites/michaeldunne/2016/12/14/chinas-automotive-2030-blueprint-no-1-globally-in-evs-autonomous-cars/#325ed6de1c6e

Eben, M. (2018). Fining Google: A missed opportunity for legal certainty? *European Competition Journal*, *14*(1), 1–23.

EC. (2017). The directive on security of network and information systems (NIS directive). European Commission. Retrieved from https://ec.europa.eu/digital-single-market/en/network-and-information-security-nis-directive

EP. (2009). European parliament resolution of 23 April 2009 on the intelligent transport systems action plan (2008/2216(INI)). Retrieved from http://www.europarl.europa.eu/sides/getDoc.do?pubRef=-//EP//TEXT+TA+P6-TA-2009-0308+0+DOC+XML+V0//EN

EP. (2017). Robots and artificial intelligence: MEPs call for EU-wide liability rules. European Parliament. Retrieved from http://www.europarl.europa.eu/news/en/press-room/20170210IPR61808/robots-and-artificial-intelligence-meps-call-for-eu-wide-liability-rules

EPCLA. (2016). Draft report with recommendations to the Commission on Civil Law Rules on Robotics (2015/2103(INL), European Parliament Committee on Legal Affairs. Retrieved from http://www.europarl.europa.eu/sides/getDoc.do?pubRef=-//EP//NONSGML+COMPARL+PE-582.443+01+DOC+PDF+V0//EN

ERTRAC. (2017). Automated driving road map. *ERTRAC Working Group "Connectivity and Automated Driving"*. Retrieved from http://www.ertrac.org/uploads/images/ERTRAC_Automated_Driving_2017.pdf

EU. (2016). GDPR key changes. European Union GDPR Portal. Retrieved from https://www.eugdpr.org/key-changes.html

Fleetwood, J. (2017). Public health, ethics, and autonomous vehicles. *American Journal of Public Health*, *107*(4), 532–537.

FMEAE. (2017). Pegasus research project. Pegasus, Federal Ministry for Economic Affairs and Energy. Retrieved from http://www.pegasus-projekt.info/en/about-PEGASUS

FMTDI. (2017a). Ethics commission on automated driving press release, Federal Ministry of Transport and Digital Infrastructure. Retrieved from https://www.bmvi.de/SharedDocs/EN/PressRelease/2017/084-ethic-commission-report-automated-driving.html

FMTDI. (2017b). Data protection law recommendations of the Federal Commissioner for Data Protection and Freedom of Information for Automated and Connected Driving. Federal Ministry of Transport and Digital Infrastructure. Retrieved from https://www.bfdi.bund.de/SharedDocs/Publikationen/Allgemein/DatenschutzrechtlicheEmpfehlungenVernetztesAuto.pdf;jsessionid=0A1EEFF8FE0D725B426F9E4E91713A3B.1_cid329?__blob=publicationFile&v=1

Frey, C. B., & Osborne, M. A. (2017). The future of employment: How susceptible are jobs to computerisation? *Technological Forecasting and Social Change*, *114*, 254–280.

Gambs, S., Killijian, M. O., & del Prado Cortez, M. N. (2014). De-anonymization attack on geolocated data. *Computer and System Sciences*, *80*(8), 1597–1614.

Gillespie, M. (2016). Shifting automotive landscapes: Privacy and the right to travel in the era of autonomous motor vehicles. *Washington University Journal of Law & Policy*, *50*, 147–169.

Glancy, D. J. (2012). Privacy in autonomous vehicles. *Santa Clara Law Review*, *52*(4), 1171–1239.

Gurney, J. K. (2013). Sue my car not me: Products liability and accidents involving autonomous vehicles. U. Ill. JL Tech. & Pol'y, 247.

Halsey, A. (2018). "We"re listening,' Department of Transportation says on the future of driverless cars. The Washington Post. Retrieved from https://www.washingtonpost.com/local/trafficandcommuting/were-listening-department-of-transportation-says-on-the-future-of-driverless-cars/2018/03/01/8992682a-1d72-11e8-b2d9-08e748f892c0_story.html?noredirect=on

Hanai, Y. (2018). Volkswagen truck teams with Toyota's Hino despite parents' rivalry. *Nikkei Asian Review*. Retrieved from https://asia.nikkei.com/Business/Companies/Volkswagen-Truck-teams-with-Toyota-s-Hino-despite-parents-rivalry

HC 143. (2017). Vehicle technology and aviation bill HC 143. *UK Parliament*. Retrieved from https://www.publications.parliament.uk/pa/bills/cbill/2016-2017/0143/cbill_2016-20170143_en_2.htm#pt1-l1g2

He, H. F. (2018). Cybersecurity law causing "mass concerns" among foreign firms in China. *South China Morning Post*. Retrieved from http://www.scmp.com/news/china/economy/article/2135338/cybersecurity-law-causing-mass-concerns-among-foreign-firms-china

Hern, A. (2016). Car hacking is the future – and sooner or later you will be hit. *The Guardian*. Retrieved from https://www.theguardian.com/technology/2016/aug/28/car-hacking-future-self-driving-security

Hevelke, A., & Nida-Rümelin, J. (2015). Responsibility for crashes of autonomous vehicles: An ethical analysis. *Science and Engineering Ethics*, *21*(3), 619–630.

Hillier, P., Wright, B., & Damen, P. (2015). Readiness for self-driving vehicles in Australia. In Workshop Report, February, ARRB Group Ltd. Retrieved from http://advi.org.au/wp-content/uploads/2016/04/Workshop-Report-Readiness-for-Self-Driving-Vehicles-in-Australia.pdf

H.R.3388. (2017). SELF DRIVE Act. Retrieved from https://www.congress.gov/bill/115th-congress/house-bill/3388/text

Japan Bullet. (2017). *New guidelines to allow driverless vehicle tests on public roads*. x: Japan Bullet http://www.japanbullet.com/news/new-guidelines-to-allowdriverless-vehicle-tests-on-public-roads

The Japan Times. (2017). Amended privacy protection law. *The Japan Times*. Retrieved from https://www.japantimes.co.jp/opinion/2017/06/01/editorials/amended-privacy-protection-law/#.WrHx-mZ7H-Z

JDSUPRA. (2017). Germany permits automated vehicles. Retrieved from http://www.jdsupra.com/legalnews/germany-permits-automated-vehicles-15610/

Jiji. (2017). Japan sets approval criteria for driverless vehicle road tests. *The Japan Times*. Retrieved from http://www.japantimes.co.jp/news/2017/06/01/business/japan-sets-approval-criteria-driverless-vehicle-road-tests/#.WVRcRcap3-Y

Kalra, N. (2017). Challenges and approaches to realising autonomous vehicle safety. Testimony submitted to the House Energy and Commerce Committee, RAND, Santa Monica, California.

Kang, C. (2016). Self-driving cars gain powerful ally: The government. *The New York Times*. Retrieved from https://www.nytimes.com/2016/09/20/technology/self-driving-cars-guidelines.html?_r=2&mtrref=tu9srvbirvvtnyr3d3cubnl0aw1lcy5jb200.g00.sandiegouniontribune.com

König, M., & Neumayr, L. (2017). Users' resistance towards radical innovations: The case of the self-driving car. *Transportation Research Part F: Traffic Psychology and Behaviour*, *44*, 42–52.

KPMG. (2017). Overview of china's cybersecurity law. *KPMG Advisory Limited*. Retrieved from https://assets.kpmg.com/content/dam/kpmg/cn/pdf/en/2017/02/overview-of-cybersecurity-law.pdf

KPMG. (2018). Autonomous vehicles readiness index. *KPMG International*. Retrieved from https://assets.kpmg.com/content/dam/kpmg/xx/pdf/2018/01/avri.pdf

Kwang, K. (2017). Changes to singapore's cybercrime law passed. *Channel News Asia*. Retrieved from http://www.channelnewsasia.com/news/singapore/changes-to-singapore-s-cybercrime-law-passed-8712368

Kyodo. (2017). NPA drafts rules for testing driverless cars on public roads. *The Japan Times*. Retrieved from http://www.japantimes.co.jp/news/2017/04/13/national/npa-drafts-rules-testing-driverless-cars-public-roads/#.WVRYVMap3-Y

Kyriakidis, M., Happee, R., & de Winter, J. C. F. (2015). Public opinion on automated driving: Results of an international questionnaire among 5000 respondents. *Transportation Research Part F: Traffic Psychology and Behaviour*, *32*, 127–140.

Lee, C. (2017). Grabbing the wheel early: Moving forward on cybersecurity and privacy protections for driverless cars. *Federal Communications Law*, *69*, 25–52.

Li, Y., Taeihagh, A., de Jong, M., & Klinke, A. (forthcoming). Reviewing and analysing the rise and spectrum of risk coping strategies – A public policy perspective.

Li, Y., Taeihagh, A., & de Jong, M. (2018). The governance of risks in ridesharing: A revelatory case from Singapore. *Energies*, *11*(5), 1277. https://doi.org/10.3390/en11051277

Liang, A., & Durbin, D. (2016). World's first self-driving taxis debut in Singapore. *Bloomberg*. Retrieved from https://www.bloomberg.com/news/articles/2016-08-25/world-s-first-self-driving-taxis-debut-in-singapore

Lim, A. (2017). Driverless vehicle rides in three new towns from 2022. *The Straits Times*. Retrieved from http://www.straitstimes.com/singapore/transport/driverless-vehicle-rides-in-three-new-towns-from-2022

Lin, P. (2016). Why ethics matters for autonomous cars. In Autonomous Driving (pp. 69-85). Springer, Berlin, Heidelberg.

Litman, T. (2017). Autonomous vehicle implementation predictions. Implications for transport planning. Victoria, Canada: Victoria Transport Policy Institute.

Long, L. N., Hanford, S. D., Janrathitikarn, O., Sinsley, G. L., & Miller, J. A. (2007, April). A review of intelligent systems software for autonomous vehicles. In *IEEE symposium on Computational intelligence in security and defence applications, CISDA, Honolulu, HI* (pp. 69–76). IEEE. Retrieved from https://ieeexplore.ieee.org/document/4219084/

Marchant, G. M., & Lindor, R. A. (2012). The coming collision between autonomous vehicles and the liability system. *Santa Clara Law Review*, *52*, 1321.

Metz, C. (2016). Artificial Intelligence is setting up the internet for a huge clash with Europe. *Wired*. Retrieved from https://www.wired.com/2016/07/artificial-intelligence-setting-internet-huge-clash-europe/

MIE. (2017). *On our way towards connected and automated driving in Europe outcome of the first high-level meeting*. Ministry of infrastructure and the environment,, Amsterdam, The Netherlands.

Milakis, D., Snelder, M., van Arem, B., Homem de Almeida Correia, G., & van Wee, G. P. (2017). Development and transport implications of automated vehicles in the Netherlands: Scenarios for 2030 and 2050. *European Journal of Transport and Infrastructure Research*, *17*(1), 63–85.

Milakis, D., van Arem, B., & van Wee, B. (2017). Policy and society related implications of automated driving: A review of literature and directions for future research. *Intelligent Transportation Systems*, *0*, 1–25.

MOT. (2017). Opening speech by second minister for transport Ng Chee Meng for the road traffic (amendment) bill second reading. Ministry of Transport, Singapore. Retrieved from https://www.mot.gov.sg/News-Centre/News/2017/Opening-Speech-by-Second-Minister-for-Transport-Ng-Chee-Meng-for-the-Road-Traffic-Amendment-Bill-Second-Reading/

MVMA 2016. Motor vehicle management Act. *Statutes of the Republic of Korea*. Retrieved from http://elaw.klri.re.kr/eng_service/lawView.do?hseq=35841&lang=ENG

Nair, S., & Howlett, M. (2016). From robustness to resilience: Avoiding policy traps in the long term. *Sustainability Science*, *11*(6), 909–917.

Narayanan, A., & Shmatikov, V. (2008, May). Robust deanonymisation of large sparse datasets. In Security and Privacy, 2008. SP 2008. IEEE Symposium on (pp. 111-125). IEEE.

NHTSA (2015). *Traffic safety facts, A brief statistical summary: Critical reasons for crashes investigated in the national motor vehicle crash causation survey*. Washington, DC: National Centre for Statistics and Analysis, U.S. Department of Transportation, DOT HS 812 115

NHTSA. (2017). Automated driving systems 2.0 a vision for safety. National Highway Traffic Safety Administration, U.S. Department of Transportation.

NHTSA. (2018). *NHTSA and vehicle cybersecurity*. Washington, DC, USA: National Highway Traffic Safety Administration.

Nicola, S., Behrmann, E., & Mawad, M. (2018). It's a good thing Europe's autonomous car testing is slow. *Bloomberg*. Retrieved from https://www.bloomberg.com/news/articles/2018-03-20/it-s-a-good-thing-europe-s-autonomous-car-testing-is-slow

Nikkei. (2015). Japan working on law aimed at governing autonomous vehicles. *Nikkei Asian Review*. Retrieved from https://asia.nikkei.com/Politics-Economy/Policy-Politics/Japan-working-on-law-aimed-at-governing-autonomous-vehicles

Nikkei. (2017). Japan looks to self-driving cars to bolster transport access. *Nikkei Asian Review*. Retrieved from https://asia.nikkei.com/Politics-Economy/Policy-Politics/Japan-looks-to-self-driving-cars-to-bolster-transport-access

Nikkei. (2018). Japan eyes black boxes for automated vehicles. Nikkei Asian Review. Retrieved from https://asia.nikkei.com/Politics/Japan-eyes-black-boxes-for-automated-vehicles

NTC. (2016). Regulatory reforms for automated vehicles *Policy Paper*, National Transport Commission.

NTC. (2017a). *National transport commission submission to the standing committee on industry, innovation, science and resources inquiry into the social issues relating to land-based driverless vehicles in Australia*. Melbourne: NTC.

NTC. (2017b). *Guidelines for trials of automated vehicles in Australia*. Melbourne: NTC. Retrieved from http://www.ntc.gov.au/Media/Reports/(00F4B0A0-55E9-17E7-BF15-D70F4725A938).pdf

NTC. (2017c). *Regulatory options to assure automated vehicle safety in Australia*. Melbourne: NTC. Retrieved from http://www.ntc.gov.au/Media/Reports/(6608D654-9FBD-175D-D067-AC80AECE5FB8).pdf

Nyholm, S., & Smids, J. (2016). The ethics of accident-algorithms for self-driving cars: An applied trolley problem? *Ethical Theory and Moral Practice*, *19*(5), 1275–1289.

Osipova, E., & Eriksson, P. E. (2013). Balancing control and flexibility in joint risk management: Lessons learned from two construction projects. *International Journal of Project Management*, *31*(3), 391–399.

Page, F. D., & Krayem, N. M. (2017). Are you ready for self-driving vehicles? *Intellectual Property & Technology Law Journal*, *29*(4), 14.

PDPC. (2018). Response to Feedback on the Public Consultation on approaches to managing personal data in the Digital Economy. Personal Data Protection Commission Singapore. Retrieved from https://www.pdpc.gov.sg/-/media/Files/PDPC/PDF-Files/Legislation-and-Guidelines/PDPC-Response-to-Feedback-for-Public-Consultation-on-Approaches-to-Managing-Personal-Data-in-the-Dig.pdf

PennDOT. (2016). Pennsylvania autonomous vehicle testing policy: Final draft report of the autonomous vehicle policy task force. Pennsylvania Department of Transportation. Retrieved from http://www.penndot.gov/ProjectAndPrograms/ResearchandTesting/Documents/AV%20Testing%20Policy%20DRAFT%20FINAL%20REPORT.pdf

Piao, J., & McDonald, M. (2008). Advanced driver assistance systems from autonomous to cooperative approach. *Transport Reviews*, *28*(5), 659–684.

Pillath, S. (2016). Automated vehicles in the EU. *EPRS, European Parliamentary Research Service, Members' Research Service*, PE 573.902, 2–12. Retrieved from http://www.europarl.europa.eu/RegData/etudes/BRIE/2016/573902/EPRS_BRI(2016)573902_EN.pdf

Pineau, J., Montemerlo, M., Pollack, M., Roy, N., & Thrun, S. (2003). Towards robotic assistants in nursing homes: Challenges and results. *Robotics and Autonomous Systems*, *42*(3), 271–281.

Pinsent Masons. (2016). Connected and autonomous Vehicles: The emerging legal challenges. Retrieved from https://www.pinsentmasons.com/en/media/publications/connected-and-autonomous-vehicles-the-emerging-legal-challenges/

PSGB. (2017). The public sector (governance) bill. Retrieved from https://sso.agc.gov.sg/Bills-Supp/45-2017/Published/20171106?DocDate=20171106

Ramirez, E. (2017). How South Korea plans to put driverless cars on the road by 2020. *Forbes*. Retrieved from https://www.forbes.com/sites/elaineramirez/2017/02/07/how-south-korea-plans-to-put-driverless-cars-on-the-road-by-2020/

Renn, O., & Benighaus, C. (2013). Perception of technological risk: Insights from research and lessons for risk communication and management. *Journal of Risk Research*, *16*(3-4), 293–313.

Reuters. (2017). Trump administration reevaluating self-driving car guidance. Fortune. Retrieved from http://fortune.com/2017/02/26/trump-self-driving-car-guidance/

RTAB. (2017). Road traffic (amendment) bill. Retrieved from http://statutes.agc.gov.sg/aol/search/display/view.w3p;ident=0c04aa1f-50dd-4078-987d-c6733dd67ec8;page=0;query=DocId%3A9fd1d504-52ec-4bf0-bb3b-3bc39a551a85%20Depth%3A0%20ValidTime%3A10%2F01%2F2017%20TransactionTime%3A10%2F01%2F2017%20Status%3Apublished;rec=0

SAE. (2014). International standard J3016 taxonomy and definitions for terms related to on-road motor vehicle automated driving systems. Retrieved from https://www.sae.org/misc/pdfs/automated_driving.pdf

SCA. (2017). S. 680, SPY car Act of 2017, 115th United States Congress. Retrieved from https://www.congress.gov/bill/115th-congress/senate-bill/680

Schellekens, M. (2016). Car hacking: Navigating the regulatory landscape. *Computer Law & Security Review: The International Journal of Technology Law and Practice*, *32*(2), 307–315.

Schoonmaker, J. (2016). Proactive privacy for a driverless age. *Information & Communications Technology Law*, *25*(2), 96–128.

SCIISR. (2017). Social issues relating to land-based automated vehicles in Australia. *Parliament of the Commonwealth of Australia*. Canberra. Retrieved from http://www.aph.gov.au/Parliamentary_Business/Committees/House/Industry_Innovation_Science_and_Resources/Driverless_vehicles/Report

Smith, B. W. (2013). Human error as a cause of vehicle crashes. *Centre for Internet and Society*. Retrieved from http://cyberlaw.stanford.edu/blog/2013/12/human-error-cause-vehicle-crashes

Spieser, K., Treleaven, K., Zhang, R. M., Frazzoli, E., Morton, D., & Pavone, M. (2014). Toward a systematic approach to the design and evaluation of automated mobility-on-demand systems: A case study in Singapore. In G. Meyer, & S Beiker (Eds.), *Road vehicle automation* (pp. 229–245). Cham: Springer.

Srikanthan, T. (2017). Commentary: Cybersecurity is the next economic battleground. Retrieved from http://www.channelnewsasia.com/news/singapore/commentary-cybersecurity-is-the-next-economic-battleground-8591642

Stahl, B. C., & Coeckelbergh, M. (2016). Ethics of healthcare robotics: Towards responsible research and innovation. *Robotics and Autonomous Systems*, *86*, 152–161.

Stone, B. (2018). Regulators are asleep at the wheel on self-driving cars. *Bloomberg*. Retrieved from https://www.bloomberg.com/news/articles/2018-03-26/regulators-are-asleep-at-the-wheel-on-self-driving-cars

The Straits Times. (2018). China completes first draft of national rules to allow road tests for driverless vehicles. *The Straits Times*. Retrieved from http://www.straitstimes.com/asia/east-asia/china-completes-first-draft-of-national-rules-to-allow-road-tests-for-driverless

Sukman, D. (2015). Lethal autonomous systems and the future of warfare. *Canadian Military*, *16*(1), 44–53.

Sun, Y., Olaru, D., Smith, B., Greaves, S., & Collins, A. (2016). Road to autonomous vehicles in Australia. Australasian transport research forum 2016, Melbourne, 16–18 November.

Teoh, E. R., & Kidd, D. G. (2017). Rage against the machine? Google's self-driving cars versus human drivers. *Journal of Safety Research*, *63*, 57–60. doi:10.1016/j.jsr.2017.08.008

Ticoll, D. (2015). Driving changes: Automated vehicles in Toronto. *Discussion paper*. Innovation Policy Lab/University of Toronto. Retrieved from https://www1.toronto.ca/City%20Of%20Toronto/Transportation%20Services/TS%20Publications/Reports/Driving%20Changes%20Final%20(compressed).pdf

Tien, J. M. (2017). The sputnik of servgoods: Autonomous vehicles. *Systems Science and Systems Engineering*, *26*(2), 133–162.

Wacket, M., Escritt, T., & Davis, T. (2017). Germany adopts self-driving vehicles law. *Reuters*. Retrieved from https://www.reuters.com/article/us-germany-autos-self-driving-idUSKBN1881HY

Wadud, Z., MacKenzie, D., & Leiby, P. (2016). Help or hindrance? The travel, energy and carbon impacts of highly automated vehicles. *Transportation Research Part A: Policy and Practice*, *86*, 1–18.

Walker, W. E., Lempert, R. J., & Kwakkel, J. H. (2013). Deep uncertainty. In S. Gass, & M. Fu (Eds.), *Encyclopaedia of operations research and management science* (pp. 395–402). New York: Springer.

West, D. M. (2016). Moving forward: Self-driving vehicles in China, Europe, Japan, Korea, and the United States. Retrieved from https://www.brookings.edu/research/moving-forward-self-driving-vehicles-in-china-europe-japan-korea-and-the-united-states/

Wildavsky, A. (1991). *Searching for safety*. New Brunswick, NJ: Transaction Books.